\documentstyle[prl,aps,epsf,floats]{revtex}

\begin{document}
\advance\textheight by 0.2in
\draft
\twocolumn[\hsize\textwidth\columnwidth\hsize\csname@twocolumnfalse\endcsname

\title{Semicircle: An exact relation in the Integer and Fractional 
Quantum Hall Effect}

\author{M. Hilke, D. Shahar$^1$, S.H. Song$^2$, D.C. Tsui, 
Y.H. Xie$^3$ and M. Shayegan}
\address{Dpt. of Elect. Eng., Princeton University, Princeton, New 
Jersey, 08544\\
$^1$Present address: Dept. of Condensed Matter Physics, Weizmann 
Institute, 
Rehovot 76100, Israel\\
$^2$Present address: LG Semicon, Choongbuk, 361-480, Korea\\
$^3$Bell Laboratories, Lucent Technologies, Murray Hill,  New 
Jersey, 07974}

\date{\today}

\maketitle
\begin{abstract}
We present experimental results on the quantized Hall insulator in 
two dimensions. This insulator, with vanishing conductivities, 
is characterized by the quantization (within experimental accuracy) of 
the Hall resistance in units of the quantum unit of resistance, $h/e^2$. 
The measurements were performed in a two dimensional hole system, confined in 
a Ge/SiGe quantum well, when the magnetic field is increased above the 
$\nu=1$ quantum Hall state. This quantization 
leads to a nearly perfect semi-circle relation
for the  diagonal and Hall conductivities.
Similar results are obtained with a higher mobility n-type modulation doped 
GaAs/AlGaAs sample, when the magnetic 
field is increased above the $\nu=1/3$ fractional quantum Hall state.
\end{abstract}
]


In the extreme quantum limit, when the magnetic field (B) exceeds a critical 
field $B_c$, the quantum Hall series is terminated by an insulator. This 
insulator is characterized by a diverging diagonal resistivity, $\rho_{xx}$, 
as the temperature (T) vanishes. In general 
this insulating behavior 
occurs when the lowest resolved energy 
level exceeds the Fermi energy.  For samples exhibiting only integer plateaus the 
transition occurs typically beyond $\nu=1$.\cite{paalanen}
For higher mobility 
samples this transition can occur beyond the $\nu=1/3$ state \cite{tsui} 
or the $\nu=1/5$ state.\cite{jiang} In addition to these primary fractions 
transitions to insulating behavior have been observed originating from 
many other fractions.\cite{sarma} In this article we will only 
concentrate on the transitions from the states $\nu=1$ to insulator and 
$\nu=1/3$ to insulator. 
For both these transitions it 
has been shown previously that 
the transition point can be obtained from the {\em B}-field, 
$B=B_c$, where a {\em T}-independent $\rho_{xx}$ is observed.\cite{shahar} 
Our main focus here is the 
determination of $\rho_{xy}$ and the inter-relation between the diagonal, 
$\sigma_{xx}$, and Hall, $\sigma_{xy}$ conductivities around the 
transitions, with emphasis on extending previous works to deep into 
the insulating phase.

In order to measure $\rho_{xy}$ we average over both {\em B-}field directions, 
which minimizes the contribution of the diverging $\rho_{xx}$. In fig. 1 we have 
plotted $\rho_{xx}$ and $\rho_{xy}$ as a function of {\em B}. The 
transition point 
$B_c=6.06$ T is easily identified with the crossing of the $\rho_{xx}$ traces
obtained at different {\em T }'s. The remarkable feature in this figure is 
the accuracy of the quantization beyond $B_c$, i.e., 
deep inside the insulating phase. The deviation 
from its quantized value between 6 and 8 T 
(inside the insulator) is less than 0.5\% for $T$=1.8 K. 
Over the entire {\em T} range (below 2 K) $\rho_{xy}$ 
deviates by less than 2\% from $h/e^2$, while $\rho_{xx}$ is highly 
insulating.

\input epsf
\begin{figure}
\epsfysize=2.5in
\epsfbox{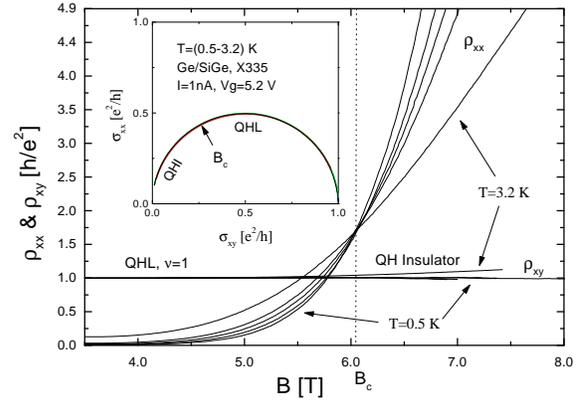}
\caption{
 The Hall and diagonal resistivities as a function 
of {\em B} for different {\em T} 's, which are 0.5, 0.8, 
1.2, 1.8 and 3.2 K. 
In the inset we have plotted $\sigma_{xx}$  as a function of $\sigma_{xy}$ for 
the same {\em T} 's. $B_c$= 6.06 T.
}
\end{figure}

Following Dykhne and Ruzin's \cite{ruzin} calculation for 
a two phase model, where the transition 
region is described by a semi-circle relation between $\sigma_{xx}$ and $\sigma_{xy}$,
we plotted in the inset of fig. 1 $\sigma_{xx}$ as a function of $\sigma_{xy}$. In their 
calculation $\sigma_{xx}^2+(\sigma_{xy}-\sigma_0/2)^2=(\sigma_0/2)^2$, where 
$\sigma_0=e^2/h$ for the $\nu=1$ to insulator transition and $\sigma_0=e^2/3h$ for 
the $\nu=1/3$ to insulator transition. Our experimental plot is close to a  
perfect semi-circle centered around $\sigma_{xy}=e^2/2h$ with radius $e^2/2h$, 
independent 
of {\em T}. These results were obtained in a two-dimensional hole gas confined in 
a Ge/SiGe quantum well, described in more details in ref. \cite{hilke}. More 
details on the {\em T} and low {\em B}-field dependences can be found 
in ref. \cite{hilke2}. 

We repeated these measurements with an n-type 
modulation doped \linebreak
GaAs/AlGaAs sample, which exhibits a 
transition from the $\nu=1/3$ quantum Hall state to insulator. We obtain  
similar results for this system, as demonstrated in fig. 2, but with a semi-circle 
centered at 0.345 $e^2/2h$, instead of 0.333 $e^2/2h$ as predicted in ref. \cite{ruzin}.
By increasing {\em T} (corresponding to the dotted lines) deviations from the semi-circle 
are becoming significant. In the inset we have plotted the corresponding 
$\rho_{xx}$ and $\rho_{xy}$ traces. Here again $\rho_{xy}$ remains within 6 \% of its 
quantized value $3h/e^2$, for the lowest measured {\em T} (0.3 K). For higher 
{\em T }'s (dotted lines), the deviations are larger. 

\begin{figure}
\epsfysize=2.5in
\epsfbox{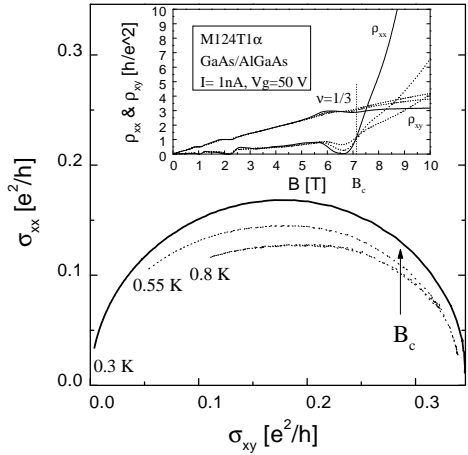}
\caption{The solid line presents $\sigma_{xx}$  as a function of $\sigma_{xy}$ 
at 0.3 K and 
the dotted lines are at 0.55 and 0.8 K. In the inset we have plotted the 
corresponding $\rho_{xx}$ 
and $\rho_{xy}$ versus {\em B}. $B_c=7.2$ T.}

\end{figure}

The quantization of $\rho_{xy}$ is intimately related to the existence of the 
semi-circle. Indeed, when obtaining conductivities from resistivities by 
matrix inversion, where $\rho_{xy}=h/\nu e^2$, we immediately obtain 
$\sigma_{xx}^2+(\sigma_{xy}-\nu e^2/2h)^2=(\nu e^2/2h)^2$, independently of 
$\rho_{xx}$. 

There has been several theoretical attempts to estimate $\rho_{xy}$ in the insulating 
phase. First Viehweger and Efetov \cite{efetov} calculated the reactive part 
of $\rho_{xy}$ and obtained $\rho_{xy}\sim B$. When calculating the diffusive part 
of $\rho_{xy}$ a finite $\rho_{xy}$ was also obtained \cite{KLZ}. More recently 
Shimshoni and Auerbach \cite{efrat}, inspired by a recent experimental work 
on duality \cite{duality}, used a semi-classical network model and 
obtained a quantized $\rho_{xy}$ in the insulating phase. 
The origin of the quantum decoherence in their model 
is, however, not clear. In a purely quantum mechanical numerical calculation in 
the lowest Landau level, i.e., for the $\nu=1$ case, a semi-circle and 
a quantized $\rho_{xy}$ was obtained very recently.\cite{bhatt}

An important point is the robustness of the quantization on 
different parameters. We therefore made measurements with different {\em T }'s, 
currents (I), contacts, density and samples. The results are summarized in 
the table. The deviation from the quantized value is obtained from the 
data for {\em B}-fields as deep as we could reliably measure inside 
the insulating phase.

\begin{center}
\footnotesize
\begin{tabular}{|c|c|}\hline
Properties for $B>B_c$ (X335) & $\rho_{xy}$ [$h/e^2$] \\ \hline
T (40 mK - 2 K) & 1 $\pm$ 2\% \\ \hline
I (0.1 nA - 1 $\mu$A) & 1 $\pm$ 5\% \\ \hline
Density (0.75-1.5) $\times 10^{11}$ cm$^{-2}$\ & 1 $\pm$ 2\% \\ \hline
Exchanging I-V contacts & 1 $\pm$ 5\% \\ \hline
Different sample (X334) & 1 $\pm$ 4\% \\ \hline
FQHE $\nu=1/3$ (M124) & 3 $\pm$ 6\% \\ \hline
\end{tabular}
\end{center}

Summarizing, we have experimentally analyzed the nature of 
a quantized Hall insulator formed when the {\em B}-field is 
increased above the $\nu=1$ (or $\nu=1/3$) to insulator transition, 
resulting in $\rho_{xy}=e^2/h\pm 2$ \% (or $\rho_{xy}=3e^2/h\pm 6\mbox{ }\%$).
This verifies the semi-circle relation. 
The experiments presented covered a certain 
range of parameters, in particular in temperature, current, density 
and disorder. Different behaviors outside of the considered range are 
however   possible and would 
be interesting to investigate in future work.

We thank A. Auerbach and L.P. Pryadko for discussions and especially S.L. Sondhi 
and E. Shimshoni for encouraging us to look at the behavior of $\rho_{xy}$ inside 
the insulator. This work was supported in part by the National 
Science Foundation and MH was supported by the Swiss National Science 
Foundation.

\end{document}